\documentclass[letterpaper, 10pt, conference]{ieeeconf}  

\IEEEoverridecommandlockouts 
\usepackage{cite}
\usepackage{amsmath,amssymb,amsfonts}
\usepackage{algorithm}
\usepackage[noend]{algpseudocode}
\usepackage{graphicx}
\usepackage{textcomp}
\usepackage{mathtools}
\usepackage{graphicx}
\usepackage{subcaption}
\usepackage{cite}
\usepackage{gensymb}
\usepackage{accents}
\usepackage{nicefrac}
\usepackage{xcolor}
\usepackage{tikz}
\usetikzlibrary{arrows.meta,calc}

\newcommand{\ubar}[1]{\underaccent{\bar}{#1}}

\newtheorem{theorem}{Theorem}[section]
\newtheorem{lemma}[theorem]{Lemma}
\newtheorem{proposition}[theorem]{Proposition}

\newtheorem{definition}[theorem]{Definition}
\newtheorem{remark}[theorem]{Remark}

\newtheorem{problem}[theorem]{Problem}
\newtheorem{example}[theorem]{Example}

\algrenewcommand{\Require}{\Statex \textbf{Input:}}  
\algrenewcommand{\Ensure}{\Statex \textbf{Output:}}

\title{Control Barrier Functions for Counting Tasks in Multi-Agent Systems}

\begin{document}

\author{
    Georgios Mitsos$^{}$,
    Sofie Haesaert$^{}$,
    Siyuan Liu$^{}$
    \thanks{This work is supported by the Dutch NWO Veni project STAMP (23061) and the European project COVER (101086228).
        The authors are with the Control Systems group in the Department of Electrical Engineering at the  Eindhoven University of Technology, the Netherlands. Emails: \{\texttt{g.mitsos, s.haesaert, s.liu5}\}\texttt{@tue.nl}
    }  
}

\maketitle

\begin{abstract}
    Multi-agent systems increasingly operate in dynamic environments and execute complex tasks that require coordination, posing significant computational challenges. 
    \emph{Counting tasks}, which require a minimum number of agents to satisfy a given condition, arise naturally in such systems. Examples include surveillance tasks, where at least $m$ drones must continuously monitor a critical area, and emergency-response tasks, where at least $m$ robots must converge to an incident location. The identities of the agents assigned to these tasks may change dynamically depending on their states, input bounds, and individual objectives.
    In this work, we propose a control barrier function (CBF)-based framework for enforcing \emph{counting invariance} and \emph{counting convergence} tasks, without explicitly assigning agents or enumerating all possible subteams.
    We develop a recursion-based method for constructing differentiable CBFs for counting tasks.
    The resulting CBF constraints are incorporated into quadratic programs, enabling online controller synthesis. 
    Numerical experiments demonstrate scalability to large multi-agent systems, while a multi-drone emergency-response case study illustrates the satisfaction of counting tasks.
\end{abstract}

\section{Introduction} \label{sec:intro}

Multi-agent systems often execute tasks that depend on the number of agents satisfying a condition rather than on their identities. Examples include smart warehouses, where at least $m$ drones must continuously patrol storage areas; disaster response, where at least $m$ robots must converge to a wildfire perimeter or earthquake impact zone; and energy systems, where a minimum number of backup generators must remain within an operating range to provide reserve capacity. Such \emph{counting tasks} can be satisfied by different agent subteams, making task allocation critical. However, preassigning agents to tasks may lead to inefficient allocations or infeasibility as operating conditions change. This motivates online controllers that dynamically adapt task allocation based on agents' states, control capabilities, and individual objectives. 

\emph{Temporal logics} provide an expressive framework for specifying high-level control objectives, including safety, reachability, and sequential or repeated execution of tasks. 
For multi-agent systems, temporal logics have been used to describe collective behaviors and synthesize
controllers with formal guarantees of task satisfaction. In \cite{kloetzer2007}, a hierarchical framework produces provably correct control laws for temporal-logic specifications over the position and size of robotic swarms. 
For counting tasks, \cite{sahin2017} introduced counting linear temporal logic (cLTL) and used integer programming to synthesize satisfying plans. This framework was extended in \cite{sahin2019} to handle bounded asynchronous execution among agents. More recently, \cite{qiu2026} developed a dual-tree-based value iteration framework to reduce computational cost for cLTL tasks in stochastic homogeneous multi-agent systems. These counting-task approaches rely on discrete system representations and offline synthesis, motivating the development of online methods operating directly in continuous state spaces.

Control barrier functions (CBFs) provide a framework for real-time control synthesis with formal safety guarantees \cite{ames2014, ames2019}. More recently, CBFs have also been used to enforce temporal-logic specifications \cite{lindemann2019, srinivasan2021} and security hyperproperties \cite{mitsos2026}. 
Boolean combinations of safety constraints can also be addressed through nonsmooth \cite{glotfelter2021} and smooth \cite{molnar2023} CBF compositions. However, a direct use of these composition methods to express counting tasks requires enumerating a combinatorial number of possible agent subteams, increasing the size of the specification and the computational cost of constructing and evaluating the associated CBF. This motivates the development of CBF formulations tailored to counting tasks.
Recently, \cite{ong2025} introduced combinatorial CBFs for \emph{counting invariance}, requiring at least $m$ out of $M$ agents to remain in a prescribed set at all times. This approach preserves the exact counting safe set and avoids combinatorial enumeration, but imposes one CBF constraint per agent. Enforcing these constraints simultaneously in a quadratic program (QP) can increase computational cost for large teams and pose significant QP feasibility challenges. 

In this paper, we develop a CBF-based framework for \emph{counting invariance} and \emph{counting convergence} tasks for large multi-agent systems. 
While counting invariance requires at least $m$ of $M$ agents to satisfy a prescribed condition at all times, counting convergence requires the system to converge to a configuration in which at least $m$ out of $M$ agents satisfy such a condition.
We associate each agent with a scalar value that is nonnegative if and only if the agent satisfies the prescribed condition. The counting task is therefore equivalent to nonnegativity of the $m^{\mathrm{th}}$ largest value.
Since this ordered value is generally nondifferentiable, we construct arbitrarily tight, differentiable lower approximations as candidate CBFs for counting tasks.
We also provide a method that efficiently evaluates these CBFs through a forward recursion and their gradients through a backward recursion, without explicit task allocation or combinatorial enumeration of agent subteams.
Compared with \cite{ong2025}, our framework handles both counting invariance and counting convergence tasks, using one aggregate CBF constraint per task rather than $M$ agent-wise constraints. We also establish sufficient conditions for locally Lipschitz continuity of the QP controller under standard feasibility and regularity assumptions.

The remainder of the paper is organized as follows. Sec.~\ref{sec:preliminaries} introduces the preliminaries and problem formulation, and Sec.~\ref{sec:cbf} constructs candidate CBFs for counting tasks. Sec.~\ref{sec:control} presents the control synthesis framework with guarantees of counting-task satisfaction. Sec.~\ref{sec:numerical} evaluates the scalability of the proposed method and presents a multi-drone case study, and Sec.~\ref{sec:conclusion} concludes the paper.
\section{Preliminaries \& Problem Formulation} \label{sec:preliminaries}

\textbf{Notation.} Let $\mathbb{R}$ and $\mathbb{R}_{\ge 0}$ denote the sets of real and nonnegative real numbers, respectively.
The zero vector is denoted $\mathbf{0}$.
For a vector $\mathbf{x} \in \mathbb{R}^n$, $\|\mathbf{x}\|$ denotes its Euclidean norm. Given a nonempty set $\mathcal{C} \subset \mathbb{R}^n$, $\|\mathbf{x}\|_\mathcal{C} \coloneqq \underset{\mathbf{y} \in \mathcal{C}}{\inf} \| \mathbf{x} - \mathbf{y} \|$ denotes the Euclidean distance of $\mathbf{x}$ from $\mathcal{C}$.
The cardinality of a set $\mathcal{I}$ is denoted by $|\mathcal{I}|$. 
For integers $M \!\ge\! 0$ and $0 \!\le \!k \!\le\! M$, the binomial
coefficient $\binom{M}{k} \coloneqq \frac{M!}{k!(M-k)!}$
denotes the number of $k$-element subsets of an $M$-element
set, with $0! \! \coloneqq \!1$. 
A continuous function $\alpha: \mathbb{R}_{\ge 0} \to \mathbb{R}_{\ge 0}$ is class-$\mathcal{K}$ if it is strictly increasing and $\alpha(0) \!= \!0$. The definition extends to extended class-$\mathcal{K}$ functions $\alpha: \mathbb{R} \to \mathbb{R}$.
The notations $\exp(x)$ and $e^x$ are used interchangeably for the exponential function.
Continuously differentiable functions are denoted by $C^1$. 

\subsection{System model}

Consider a multi-agent system with $M$ agents, indexed by $\mathcal{I} = \{1, \dots, M\}$. Each agent~$i \in \mathcal{I}$ follows nonlinear input-affine dynamics, i.e.,
\begin{equation} \label{eq:system}
    \dot{\mathbf{x}}_i = \mathbf{f}_i(\mathbf{x}_i) + \mathbf{G}_i(\mathbf{x}_i) \, \mathbf{u}_i, \quad i \in \mathcal{I},
\end{equation}
where $\mathbf{x}_i \in \mathbb{R}^n$ and $\mathbf{u}_i \in \mathbb{R}^p$ are its state and control input, respectively. The functions $\mathbf{f}_i: \mathbb{R}^n \rightarrow \mathbb{R}^n$ and $\mathbf{G}_i: \mathbb{R}^n \rightarrow \mathbb{R}^{n \times p}$, $i \in \mathcal{I}$, are locally Lipschitz continuous. Let
\begin{equation} \label{eq:joint_system}
    \widetilde{\mathbf{x}} \!=\! \begin{bmatrix} \mathbf{x}_1^\top & \!\!\!\! \dots \!\!\!\! & \mathbf{x}_M^\top \end{bmatrix}^\top \!\! \in \mathbb{R}^{M n}, 
    \widetilde{\mathbf{u}} \!=\! \begin{bmatrix} \mathbf{u}_1^\top & \!\!\!\! \dots \!\!\!\! & \mathbf{u}_M^\top \end{bmatrix}^\top \!\! \in \mathbb{R}^{M p}
\end{equation}
denote the stacked state and input, respectively. Note that given a locally Lipschitz controller, the closed-loop system is locally Lipschitz. Then, for any initial condition $\widetilde{\mathbf{x}}(0) = \widetilde{\mathbf{x}}_0$, there exists a unique maximal solution $\widetilde{\mathbf{x}}(t)$ defined on $I \left(\widetilde{\mathbf{x}}_0 \right) = [0, \tau_{\max})$. The closed-loop system is forward complete if $\tau_{\max} = \infty$ for any $\widetilde{\mathbf{x}}_0$.

\subsection{Control barrier functions}

Let $\widetilde{\mathcal{D}} \subseteq \mathbb{R}^{Mn}$ be an open set, and define the set $\widetilde{\mathcal{C}} \subset \widetilde{\mathcal{D}}$ as the zero-superlevel set of a $C^1$ function $H : \widetilde{\mathcal{D}} \to \mathbb{R}$, i.e., $\widetilde{\mathcal{C}} \coloneqq \left\{ \widetilde{\mathbf{x}} \in \widetilde{\mathcal{D}} : H(\widetilde{\mathbf{x}}) \ge 0 \right\} $ .

\begin{definition} \label{def:cbf}
    A $C^1$ function $H : \widetilde{\mathcal{D}} \to \mathbb{R}$ is a control barrier function (CBF) for $\widetilde{\mathcal{C}}$ if there exists a locally Lipschitz extended class-$\mathcal{K}$ function $\alpha$ such that, for all $\widetilde{\mathbf{x}} \in \widetilde{\mathcal{D}}$, 
    \begin{equation} \label{eq:cbf}
        \sup_{\substack{\mathbf{u}_i \in \mathbb{R}^p \\ i \in \mathcal{I}}} \,
        \sum_{i=1}^{M} \frac{\partial H}{\partial\mathbf{x}_i} (\widetilde{\mathbf{x}}) \left( \mathbf{f}_i (\mathbf{x}_i) + \mathbf{G}_i (\mathbf{x}_i) \, \mathbf{u}_i \right) \ge -\alpha(H(\widetilde{\mathbf{x}})).
    \end{equation}
\end{definition} 

\vspace{0.5em}
Note that Def.~\ref{def:cbf} implies a centralized control framework, that is, each agent's control input $\mathbf{u}_i$ may depend on the joint state $\widetilde{\mathbf{x}}$. This definition is used in Sec.~\ref{sec:control} to synthesize online controllers for counting tasks.

\subsection{Problem formulation}

Consider the system in \eqref{eq:system}, and $m_\mathrm{inv}, m_\mathrm{conv} \in \{1, \dots, M\}$. Consider the $C^1$ functions $h_\mathrm{inv}, h_\mathrm{conv}: \mathbb{R}^n \to \mathbb{R}$, and the sets
\begin{equation*}
    \mathcal{C}_\mathrm{inv} = \{ \mathbf{x} : h_\mathrm{inv}(\mathbf{x}) \ge 0 \}, \
    \mathcal{C}_\mathrm{conv} = \{ \mathbf{x} : h_\mathrm{conv}(\mathbf{x}) \ge 0 \} \subset \mathbb{R}^n.
\end{equation*}
Define, for any set $\mathcal{C}$, $\mathbb{I}_{\mathcal{C}}(\mathbf{x}) = 1$ if $\mathbf{x} \in \mathcal{C}$, and 0 otherwise.
\begin{definition} \label{def:counting_invariance}
     A \emph{counting invariance task} is satisfied if at least $m_\mathrm{inv}$ out of the $M$ agents in \eqref{eq:system} belong to $\mathcal{C}_\mathrm{inv}$ at all times, i.e., $\sum_{i=1}^{M} \mathbb{I}_{\mathcal{C}_\mathrm{inv}}(\mathbf{x}_i(t)) \ge m_\mathrm{inv}$, $\forall t \in \mathbb{R}_{\ge 0}$. 
\end{definition}

The identities of the agents satisfying the counting invariance task need not be fixed over time, provided that the required number $m_\mathrm{inv}$ is maintained.
\begin{definition} \label{def:counting_convergence}
    A \emph{counting convergence task} is satisfied if the joint state $\widetilde{\mathbf{x}}$ of \eqref{eq:system} eventually converges to the set $\widetilde{\mathcal{T}}_\mathrm{conv} = \left\{ \widetilde{\mathbf{x}} \in \mathbb{R}^{M n} : \sum_{i=1}^{M} \mathbb{I}_{\mathcal{C}_\mathrm{conv}}(\mathbf{x}_i) \ge m_\mathrm{conv} \right\}$
    as $t \to \infty$.
\end{definition}

\begin{problem} \label{pr:counting_tasks}
    Design controllers $\mathbf{u}_i (\widetilde{\mathbf{x}}) \in \mathbb{R}^p$, $i \in \mathcal{I}$, such that the system in \eqref{eq:system} satisfies the counting invariance task in Def.~\ref{def:counting_invariance} and/or the counting convergence task in Def.~\ref{def:counting_convergence}. 
\end{problem}

To solve Prob.~\ref{pr:counting_tasks}, we first construct differentiable candidate CBFs for the counting tasks and develop an efficient recursion-based approach for evaluating their values and gradients (cf. Sec.~\ref{sec:cbf}). Then, we incorporate the corresponding CBF constraints into a QP that enables controller synthesis for counting invariance and convergence without explicitly assigning agents to tasks (cf. Sec.~\ref{sec:control}). 
 
\section{Candidate CBFs for Counting Tasks} \label{sec:cbf}

In this section, we develop a method for constructing and efficiently evaluating differentiable CBFs.

Consider the system in \eqref{eq:system}.
Let $h: \mathbb{R}^n \to \mathbb{R}$ be a 
$C^1$ function defining a set $\mathcal{C} = \left\{ \mathbf{x} \in \mathbb{R}^n: h(\mathbf{x}) \ge 0 \right\}$. This function is evaluated at each agent's state, 
such that $h(\mathbf{x}_i)\ge0$ if and only if agent~$i$ belongs to $\mathcal{C}$.
The construction of candidate CBFs in this section applies to both counting invariance tasks with $h = h_\mathrm{inv}$ and counting convergence tasks with $h = h_\mathrm{conv}$. 
We denote the $j^\mathrm{th}$ largest among these values by $h_{(j)}$, $j \in \mathcal{I}$, such that $h_{(1)} \ge h_{(2)} \ge \dots \ge h_{(M)}$. 
Requiring at least $m$ agents to belong to $\mathcal{C}$ is thus equivalent to requiring $h_{(m)} \ge 0$. 
Although evaluating $h_{(m)}$ is straightforward, the function is generally nondifferentiable at points where the ordering of the values changes, preventing its direct use as a CBF for counting tasks. Therefore, we seek a differentiable lower approximation $\ubar{h}_{(m)} \le  h_{(m)}$ to serve as a candidate CBF. Enforcing $\ubar{h}_{(m)} \ge 0$ guarantees $h_{(m)} \ge 0$, and hence ensures that at least $m$ agents belong to $\mathcal{C}$.

We start by expressing $h_{(m)}$ in terms of sums of ordered values. The sum of the $k$ largest values is 
\begin{equation*}
    \sigma_k = \sum_{j=1}^k h_{(j)}, \ k = 1, \dots, M,
\end{equation*}
with $\sigma_0 = 0$. We denote by $\mathcal{I}_k$ the collection of all subteams consisting of $k$ agents, i.e., $\mathcal{I}_k \coloneqq \{I \subseteq \mathcal{I}: |I| = k \}$. Then, we can express  $\sigma_k$ in terms of the agents' values $h(\mathbf{x}_i)$ as 
\begin{equation} \label{eq:sum_k_largest}
    \sigma_k = \max_{I \in \mathcal{I}_k} \sum_{i \in I} h(\mathbf{x}_i).
\end{equation}
Intuitively, among all possible subteams of $k$ agents, the largest sum $\sigma_k$ is obtained by selecting the agents associated with the $k$ largest values. The $m^\mathrm{th}$ largest value, $h_{(m)}$, can then be recovered from two consecutive sums as 
\begin{equation} \label{eq:mth_largest}
    h_{(m)} = \sigma_m - \sigma_{m-1}. 
\end{equation}
Note that $\!\sigma_m $ and $ \sigma_{m\!-\!1}$ are still nondifferentiable. Next, we construct $C^1\!$ approximations of these two sums and develop a recursion-based method to efficiently evaluate the derived candidate CBF and its gradient.

\subsection{Construction of candidate CBFs}

To obtain a $C^1$ approximation of \eqref{eq:mth_largest}, we approximate the sums $\sigma_m$ and $\sigma_{m-1}$ using the LogSumExp function.
Recall that, for any $y_1, \dots, y_N \in \mathbb{R}$ and $\tau > 0$, the following inequality holds
\begin{equation} \label{eq:lse}
    \max_i y_i \le \tau \ln \left( e^{\tfrac{y_1}{\tau}} + \dots + e^{\tfrac{y_N}{\tau}} \right) \le \max_i y_i + \tau \ln N.
\end{equation}
In order to apply it to lower approximate $\sigma_k$ in \eqref{eq:sum_k_largest}, we define
\begin{equation} \label{eq:hat_sum_k_largest}
\!\!\!\!  \hat{\sigma}_k \coloneqq \tau \ln \left( 
\sum_{I_j \in \mathcal{I}_k , j \in |\mathcal{I}_k|} e^{ \frac{y_j}{\tau} } \right)\!, \!\!
\text{ where } y_j = \sum_{i \in I_j} h(\mathbf{x}_i).
\end{equation}
For $k = 1, \dots, M$, the collection $\mathcal{I}_k$ contains $\tbinom{M}{k}$ subteams, i.e., $ |\mathcal{I}_k| = \tbinom{M}{k},$ corresponding to all ways of selecting $k$ agents of the system. Thus, applying \eqref{eq:lse} to these $\tbinom{M}{k}$ subteam sums gives $\sigma_k \le \hat{\sigma}_k \le \sigma_k + \tau \ln \tbinom{M}{k}$, or, equivalently, 
\begin{equation*}
    \hat{\sigma}_k - \tau \ln \tbinom{M}{k} \le \sigma_k \le \hat{\sigma}_k. 
\end{equation*}
We can now construct a $C^1$ lower approximation of $h_{(m)} = \sigma_m - \sigma_{m-1}$, by lower-bounding $\sigma_m$ and upper-bounding $\sigma_{m-1}$. We have
\begin{equation} \label{eq:underapprox_recursion}
    h_{(m)} = \sigma_m - \sigma_{m-1} \ge \hat{\sigma}_m - \tau \ln \tbinom{M}{m} - \hat{\sigma}_{m-1} \coloneqq \ubar{h}_{(m)}.
\end{equation}
Therefore, $\ubar{h}_{(m)} \ge 0$ implies $h_{(m)} \ge 0$, which guarantees that at least $m$ agents belong to $\mathcal{C}$. The conservatism introduced by this approximation can be explicitly bounded as
\begin{equation} \label{eq:approx_width1}
    0 \le h_{(m)} - \ubar{h}_{(m)} \le \tau \left( \ln \tbinom{M}{m} + \ln \tbinom{M}{m-1} \right).
\end{equation}
The approximation becomes arbitrarily tight as $\tau \to 0$, with $\ubar{h}_{(m)} \to h_{(m)}$. We have now obtained a differentiable lower approximation of $h_{(m)}$, i.e., $\ubar{h}_{(m)}$, which serves as a candidate CBF. However, directly evaluating $\hat{\sigma}_m$ and $\hat{\sigma}_{m-1}$ requires enumerating a combinatorial number of subteams and becomes prohibitive for online computation as the number of agents increases (cf. Fig.~\ref{fig:comparison1}). Next, we show how we can compute these quantities recursively without explicit subteam enumeration.

\subsection{Forward recursion for efficient CBF evaluation}\label{subsec:CBFrecursion}

We compute the candidate CBF by processing all agents sequentially and accumulating the exponential contributions of all possible subteams of each size as in \eqref{eq:hat_sum_k_largest}. We use \emph{stage}~$l = 0, \dots, M$ to indicate that the first  $l$ agents have been processed.
For each stage $l = 0, \dots, M$ and subteam size $k = 0, \dots, m$, let us define  
\begin{equation*} 
    Z_k^{(l)} \coloneqq \sum_{I \subseteq \{1, \dots, l \} : \, |I| = k} \exp \left({ \frac{1}{\tau} \sum_{i\in I} h(\mathbf{x}_i)} \right).
\end{equation*}
At the final stage $l = M$, all agents have been processed, thus \eqref{eq:hat_sum_k_largest} gives 
\begin{equation} \label{eq:sigma_to_Z}
    \hat{\sigma}_k = \tau \ln Z_k^{(M)}.
\end{equation}
Substituting this into \eqref{eq:underapprox_recursion} yields
\begin{equation}\label{eq:lower_approximation1}
    \ubar{h}_{(m)}=\tau\ln Z_m^{(M)}-\tau\ln Z_{m-1}^{(M)}
    -\tau\ln\tbinom{M}{m}.
\end{equation}
Therefore, evaluating the candidate CBF $\ubar{h}_{(m)}$ requires only the terminal quantities $Z_m^{(M)}$ and $Z_{m-1}^{(M)}$.

To update $Z_k^{(l)}$, we partition the $k$-agent subteams drawn from the first $l$ agents into two disjoint groups, according to whether they contain agent~$l$. 
Subteams in the first group exclude agent~$l$ and are formed by selecting $k$ agents from the first $l-1$ agents. Thus, their total contribution is $Z_k^{(l-1)}$.  
Subteams in the second group are formed by adding agent~$l$ to a $(k-1)$-agent subteam selected from the first $l-1$ agents. The total contribution of the second group is therefore $\exp\!\left(h(\mathbf{x}_l)/\tau\right)Z_{k-1}^{(l-1)}$. 
Combining the two groups, we obtain
\begin{equation} \label{eq:recursion}
Z_k^{(l)} = Z_k^{(l-1)} + \exp\left(\frac{h(\mathbf{x}_l)}{\tau}\right) Z_{k-1}^{(l-1)}.
\end{equation}

We initialize the recursion by setting $Z_0^{(l)}=1$ for $l=0,\dots,M$, since the empty subteam contributes one. We also set $Z_k^{(l)}=0$ whenever $k>l$, because a $k$-agent subteam cannot be formed from only $l < k$  agents.
At each stage $l=1,\dots,M$, we apply \eqref{eq:recursion} for $k=1,\dots,\min\{l,m\}$, using values only from stage $l-1$. This forward recursion produces the two terminal quantities required in \eqref{eq:lower_approximation1}.


\begin{figure}[t]
\centering
\begingroup
\begin{tikzpicture}[x=1.60cm,y=1.40cm,
  font=\footnotesize,line cap=round,line join=round,
  recdot/.style={circle,draw=black,fill=black,minimum size=2.2mm,inner sep=0pt},
  recend/.style={circle,draw=black,dashed,fill=black!35,minimum size=2.5mm,inner sep=0pt},
  recedge/.style={-{Latex[length=1.6mm,width=1.2mm]},draw=black,line width=.5pt}
]
\definecolor{recgreen}{RGB}{171,226,162}
\definecolor{recred}{RGB}{249,139,136}
\definecolor{recblue}{RGB}{119,111,248}
\fill[recgreen] (-.18,-.20) rectangle (2.52,.20);
\fill[recred] (-.18,-.66) -- (-.18,-3.13) -- (2.28,-3.13) -- cycle;
\fill[recblue] (2.58,.20) -- (4.15,.20) -- (4.15,-1.38) -- cycle;

\foreach \stage in {0,...,4}{
  \node at (\stage,.52) {\(l=\stage\)};
}
\foreach \card in {0,...,3}{
  \node[anchor=east] at (-.24,-\card) {\(k=\card\)};
}

\foreach \stage in {0,...,4}{
  \foreach \card in {0,...,3}{
    \ifnum\stage=4
      \ifnum\card>1
        \node[recend] (rn-\stage-\card) at (\stage,-\card) {};
      \else
        \node[recdot] (rn-\stage-\card) at (\stage,-\card) {};
      \fi
    \else
      \node[recdot] (rn-\stage-\card) at (\stage,-\card) {};
    \fi
    \ifnum\card=0
      \def\reclabely{0}
      \ifnum\stage=4
        \def\reclabelx{.18}
      \else
        \def\reclabelx{.11}
      \fi
    \else
      \def\reclabelx{.025}
      \def\reclabely{.20}
      \ifnum\stage=4
        \ifnum\card=1 \def\reclabelx{.18}\fi
      \fi
    \fi
    \def\reclabelmath{Z_{\card}^{(\stage)}}
    \ifnum\stage=4
      \ifnum\card>1
        \def\reclabelmath{\boldsymbol{Z_{\card}^{(\stage)}}}
      \fi
    \fi
    \node[anchor=west,inner sep=.3pt]
      at ($(rn-\stage-\card.center)+(\reclabelx,\reclabely)$)
      {\(\reclabelmath\)};
  }
}

\foreach \src/\dst in {
  0-1/1-1,1-1/2-1,2-1/3-1,
  1-2/2-2,2-2/3-2,3-2/4-2,
  2-3/3-3,3-3/4-3}{
  \draw[recedge] (rn-\src.east) -- (rn-\dst.west);
}

\foreach \src/\dst/\agent in {
  0-0/1-1/1,1-0/2-1/2,2-0/3-1/3,
  1-1/2-2/2,2-1/3-2/3,3-1/4-2/4,
  2-2/3-3/3,3-2/4-3/4}{
  \draw[recedge] (rn-\src.south east) --
    node[pos=.60,sloped,above=1.0pt,font=\scriptsize,inner sep=.2pt]
      {\(+\)\,Agent~\agent}
    (rn-\dst.north west);
}

\end{tikzpicture}
\endgroup
\caption{Illustration of the forward recursion for $M=4$ and $m=3$. Green entries equal one, red entries equal zero, and blue entries do not need to be evaluated.
}
\label{fig:recursion}
\vspace{-0.5cm}
\end{figure}
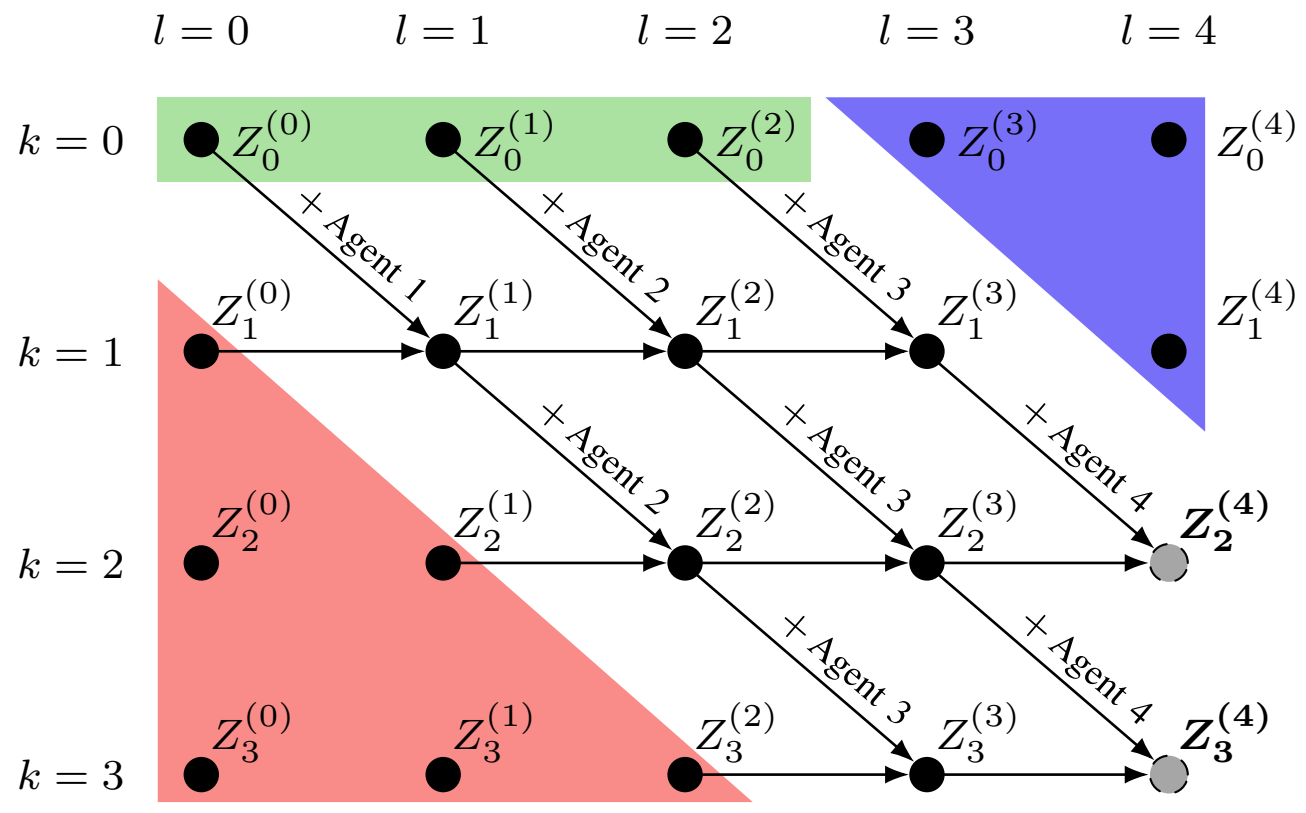 

The following example illustrates how the forward recursion evaluates the candidate CBF by updating and reusing intermediate quantities across successive stages.
\begin{example}
    Consider a system with $M=4$, $m=3$, $\tau=1$, and $h(\mathbf{x}_1) = 0.8$, $h(\mathbf{x}_2) = 0.45$, $h(\mathbf{x}_3) = -0.4$, $h(\mathbf{x}_4) = -0.2$. Fig.~\ref{fig:recursion} shows the recursion process, with stages $l$ in columns and subteam sizes $k$ in rows. Horizontal and diagonal arrows represent the two terms in \eqref{eq:recursion}. The gray nodes mark the terminal sums $Z_2^{(4)}$ and $Z_3^{(4)}$ used to evaluate the candidate CBF as in \eqref{eq:lower_approximation1}.
    
    At stage~$l=1$, the only way to select one agent is to select agent~1. Thus, $Z_1^{(1)} \!=\! e^{h(\mathbf{x}_1)} \!=\! e^{0.8}$. At stage~$l=2$, there are two ways to select one agent, either agent~1 or agent~2. Thus,    
    $ Z_1^{(2)}=Z_1^{(1)}+e^{h(\mathbf{x}_2)} = e^{0.8} + e^{0.45}.$ Moreover, the only way to select two agents is to include both agents~1 and 2. Thus,    
    $ Z_2^{(2)} = e^{h(\mathbf{x}_2)} Z_1^{(1)} = e^{0.8 + 0.45} = e^{1.25}.$
    At stage $l=3$, the two-agent subteam $\{1,2\}$ excludes agent~3, whereas $\{1,3\}$ and $\{2,3\}$ include it. Accordingly,
        \begin{align*}
        Z_2^{(3)}
        =Z_2^{(2)}+e^{-0.4}Z_1^{(2)}
          \approx6.033.
    \end{align*}
    These two contributions correspond to the two arrows entering $Z_2^{(3)}$ in Fig.~\ref{fig:recursion}. At the final stage, $Z_2^{(4)}\approx9.688$ and $Z_3^{(4)}\approx7.279$. Since the third-largest value is $h_{(3)}=h(\mathbf{x}_4)=-0.2$, \eqref{eq:lower_approximation1} yields
    \begin{align*}
        \ubar{h}_{(3)} \!= 
        \ln Z_3^{(4)} - \ln Z_2^{(4)} - \ln\! \binom{4}{3}
        \approx-1.672\le h_{(3)} \!= -0.2.
    \end{align*}
    For $\tau = 1$, the approximation is conservative, as expected from the bound in \eqref{eq:approx_width1}, which gives
    \[
        0 \le h_{(3)} - \ubar{h}_{(3)} \le 3.178.
    \]
    However, $\ubar{h}{(3)}$ remains a valid lower approximation of $h_{(3)}$. Reducing $\tau$ to $0.1$ and $0.01$ yields $\ubar{h}{(3)} \approx -0.326$ and $\approx -0.214$, respectively, showing convergence to $h_{(3)}$ as $\tau \to 0$, consistent with the approximation error bounds in \eqref{eq:approx_width1}.
    \hfill $\diamond$
\end{example}

Since only the two terminal sums are needed to evaluate the candidate CBF in \eqref{eq:lower_approximation1}, some entries of the recursion do not need to be computed.
A $k$-agent subteam at stage~$l$ can grow to at most $k+M-l$ agents by stage~$M$, so it cannot contribute to either terminal sum if $k+M-l<m-1$. Thus, the update range can be restricted to
\[
     k_\mathrm{min}^{(l)} = \max\{1,m-M+l-1\} \le k \le \min\{l,m\}.
\]
In Fig.~\ref{fig:recursion}, the blue entries can be omitted for this reason, while the red entries are zero. With this restricted update range, the constant entries $Z_0^{(l)}=1$ are needed only for $0\le l\le M-m+1$, as shown in green.

Next, we differentiate the forward recursion to efficiently evaluate the gradient required by the CBF constraint in \eqref{eq:cbf}.



\subsection{Backward recursion for CBF gradient evaluation} \label{subsec:CBFgradientrecursion}


We compute the gradient of $\ubar{h}_{(m)}$ by differentiating the forward recursion in \eqref{eq:recursion}.
We first propagate derivatives backward from the terminal quantities, then apply the chain rule to obtain the derivatives with respect to $h(\mathbf{x}_l)$ and the agent states $\mathbf{x}_l$, $l = 1, \dots, M$.


For each entry, we define $\lambda_k^{(l)} \coloneqq \nicefrac{\partial \ubar{h}_{(m)}}{\partial Z_k^{(l)}}$. 
This derivative accounts for all subsequent stages through which 
$Z_k^{(l)}$ affects the candidate CBF. 
At the final stage $l = M$, \eqref{eq:lower_approximation1} depends only on $Z_m^{(M)}$ and $Z_{m-1}^{(M)}$, so we initialize the backward recursion with 
\begin{equation*}      
    \lambda_m^{(M)}=\frac{\tau}{Z_m^{(M)}}, \quad \lambda_{m-1}^{(M)}=-\frac{\tau}{Z_{m-1}^{(M)}},
\end{equation*}
and $\lambda_k^{(M)} = 0$, for $k=0, \dots, m-2$. Starting from these terminal derivatives, we proceed backward through stages $l = M-1, \ldots, 1$. By \eqref{eq:recursion}, each $Z_k^{(l)}$ contributes to two entries at stage $l+1$: $Z_k^{(l+1)}$ and $Z_{k+1}^{(l+1)}$. Hence, the chain rule sums the contributions through these two entries: 
\begin{equation*}
    \lambda_k^{(l)} = \lambda_k^{(l+1)} + e^{\frac{h(\mathbf{x}_{l+1})}{\tau}} \lambda_{k+1}^{(l+1)}, \, k = k_\mathrm{min}^{(l)}, \dots, \min \{ l, m-1\}.
\end{equation*}
If $l\ge m$, the entry $Z_m^{(l)}$ contributes only to $Z_m^{(l+1)}$, giving $\lambda_m^{(l)}=\lambda_m^{(l+1)}$. 
As in the forward recursion, entries with $1 \le k < k_\mathrm{min}^{(l)}$ are omitted, and their derivatives are set to zero. 
The constant entries $Z_0^{(l)}=1$ and $Z_k^{(l)}=0$ for $k>l$ require no backward updates.


Once the derivatives at stage $l$ are available, we compute the derivative with respect to $h(\mathbf{x}_l)$. Note that $h(\mathbf{x}_l)$ appears explicitly in the forward recursion in \eqref{eq:recursion} only at stage $l$. Thus, differentiating \eqref{eq:recursion} gives
\begin{equation*}
    \frac{\partial Z_k^{(l)}}{\partial h(\mathbf{x}_l)} = \frac{1}{\tau} e^{\frac{h(\mathbf{x}_l)}{\tau}} Z_{k-1}^{(l-1)}, \quad k = k_\mathrm{min}^{(l)}, \dots, \min\{l, m\}.
\end{equation*}
Since $h(\mathbf{x}_l)$ affects all entries at stage $l$ with $k \ge 1$, we apply the chain rule and sum their contributions to obtain
\begin{align*}
    \frac{\partial \ubar{h}_{(m)}}{\partial h(\mathbf{x}_l)} &=\! \sum_{k = k_\mathrm{min}^{(l)}}^{\min\{l,m\}} \frac{\partial \ubar{h}_{(m)}}{\partial Z_k^{(l)}} \frac{\partial Z_k^{(l)}}{\partial h(\mathbf{x}_l)} = \\
    &=\! \sum_{k = k_\mathrm{min}^{(l)}}^{\min\{l,m\}} \! \lambda_k^{(l)} \frac{\partial Z_k^{(l)}}{\partial h(\mathbf{x}_l)} = \frac{1}{\tau} e^{\frac{h \left(\mathbf{x}_l \right)}{\tau}} \! \sum_{k = k_\mathrm{min}^{(l)}}^{\min\{l,m\}} \! \lambda_k^{(l)} Z_{k-1}^{(l-1)}.
\end{align*}
Finally, multiplying by the gradient of $h(\mathbf{x}_l)$ gives the desired state gradient:
\begin{equation} \label{eq:derivative1}
    \frac{\partial \ubar{h}_{(m)}}{\partial \mathbf{x}_l} = \frac{\partial\ubar{h}_{(m)}}{\partial h(\mathbf{x}_l)} \frac{\partial h(\mathbf{x}_l)}{\partial\mathbf{x}_l}, \quad l = 1, \dots, M.
\end{equation}

\begin{algorithm}[th!]
\caption{Recursions for CBF and gradient evaluation}
\label{alg:recursion}
\begin{algorithmic}[1]
\Require \quad $m$, $\tau$, $h(\mathbf{x}_l)$,
$\nicefrac{ \partial h(\mathbf{x}_l)}{\partial \mathbf{x}_l}$, $l = 1, \dots, M$
\Ensure \quad $\ubar{h}_{(m)}$, $\nicefrac{ \partial \ubar{h}_{(m)} }{\partial \mathbf{x}_l}$, $l = 1, \dots, M$

\State $k_\mathrm{min}^{(l)} = \max \{1, m-M+l-1\}$ 
\State Initialize the forward recursion: $Z_0^{(l)}\gets 1$, $l = 0, \dots, M$, and $Z_k^{(l)} \gets 0$, $k>l$
\For{$l = 1, \dots, M$}
    \For{$k = k_\mathrm{min}^{(l)}, \dots, \min \{ l, m \}$}
        \State $Z_k^{(l)} \gets Z_k^{(l-1)} +\exp \left(\nicefrac{h \left(\mathbf{x}_l \right)}{\tau} \right) Z_{k-1}^{(l-1)}$
    \EndFor
\EndFor

\State $\ubar{h}_{(m)} \gets \tau \ln Z_m^{(M)} - \tau \ln Z_{m-1}^{(M)} - \tau \ln \binom{M}{m}$

\vspace{1mm}
\State Initialize the backward recursion: 
\Statex $\lambda_k^{(l)} \gets 0$, $k = 0, \dots, m$, $l= 0, \dots, M$, 
\Statex $\lambda_m^{(M)} \gets \nicefrac{\tau}{Z_m^{(M)}}$, $\lambda_{m-1}^{(M)} \gets \nicefrac{-\tau}{Z_{m-1}^{(M)}}$ 

\For{$l = M, \dots, 1$}
    \If{$l<M$}
        \For{$k = k_\mathrm{min}^{(l)} , \dots, \min \{l, m-1 \}$}
            \State $\displaystyle \lambda_k^{(l)} \gets \lambda_k^{(l+1)} + \exp \left(\nicefrac{h \left( \mathbf{x}_{l+1} \right)}{\tau} \right) \lambda_{k+1}^{(l+1)}$
        \EndFor
        \If{$m \le l$}
            \State $\lambda_m^{(l)} \gets \lambda_m^{(l+1)}$
        \EndIf
    \EndIf

    \State $\displaystyle \frac{\partial \ubar{h}_{(m)}}{\partial h(\mathbf{x}_l)} \gets \frac{1}{\tau} \exp \left(\nicefrac{h \left( \mathbf{x}_l \right)}{\tau} \right) \sum_{k = k_\mathrm{min}^{(l)}}^{\min\{l,m\}} \lambda_k^{(l)} Z_{k-1}^{(l-1)}$

    \vspace{1mm}
    \State $\displaystyle
    \frac{\partial\ubar{h}_{(m)}}{\partial\mathbf{x}_l}
    \gets
    \frac{\partial\ubar{h}_{(m)}}{\partial h(\mathbf{x}_l)}
    \frac{\partial h(\mathbf{x}_l)}{\partial\mathbf{x}_l}$
\EndFor
\end{algorithmic}
\end{algorithm}



Alg.~\ref{alg:recursion} summarizes the forward and backward recursions presented in Sec.~\ref{subsec:CBFrecursion} and \ref{subsec:CBFgradientrecursion}. 
Given the individual agents' $h(\mathbf{x}_i)$ and their gradients, 
Lines~1--5 initialize and evaluate the forward recursion, producing the terminal sums required to compute the candidate CBF $\ubar{h}_{(m)}$ in line~6. 
The backward recursion in lines~7--15 reuses the stored intermediate quantities from the forward recursion to compute the gradient with respect to the agent states $\mathbf{x}_l$ (line~15).
These outputs are used in Sec.~\ref{sec:control} to formulate the CBF constraints for counting
invariance and convergence tasks.

The proposed approach constructs a $C^1$ candidate CBF for counting tasks and efficiently evaluates its value and gradient through forward and backward recursions, without explicitly enumerating agent subteams. Moreover, the approach is permutation invariant: reordering the agents leaves the CBF value unchanged and permutes its gradient components accordingly.


\subsection{Complexity analysis} \label{sec:complexity}

Next, we analyze the computational and memory complexity of the proposed method. Direct evaluation of $\hat{\sigma}_m$ and $\hat{\sigma}_{m-1}$ as in \eqref{eq:hat_sum_k_largest} requires enumerating $\binom{M}{m}$ and $\binom{M}{m-1}$ subteams, respectively. Since computing each subteam sum requires $\mathcal{O}(m)$ operations, direct enumeration has computational complexity $\mathcal{O} \left(m \!\cdot\! \tbinom{M}{m} \right)$, 
which is combinatorial in $M$ and quickly becomes prohibitive for online computation. 

The proposed recursion-based method processes at most $m$ entries over $M$ stages, with a constant number of operations per entry in both the forward and backward recursions. Thus, the computational complexity is $\mathcal{O}(M m)$. 
For $m > \nicefrac{(M+1)}{2}$, it is more efficient to view the $m^\mathrm{th}$ largest value as the $(M-m+1)^\mathrm{th}$ smallest value, and express it as the difference between the sum of the $M-m+1$ and $M-m$ smallest values. These sums can be smoothly approximated using the LogSumExp approximation of the minimum. The same forward and backward recursions then apply, with only $M-m+1$ instead of $m$ entries per stage. Thus, the computational complexity is $ \mathcal{O} \left(M \!\cdot\! \min\{m, M-m+1\} \right)$, with worst-case complexity $\mathcal{O}(M^2)$.
The backward recursion requires storing intermediate quantities from the forward recursion. Thus, the memory complexity is $\mathcal{O} \left(M \!\cdot\! \min\{m, M-m+1\} \right)$, or $\mathcal{O}(M^2)$ in the worst case. 

\section{Controller Synthesis} \label{sec:control}

In this section, we use the candidate CBFs and their gradient developed in Sec.~\ref{sec:cbf} to synthesize controllers for the counting tasks defined in Def.~\ref{def:counting_invariance}-\ref{def:counting_convergence}. 
We first establish sufficient conditions for forward invariance and attractivity of sets. We then incorporate the corresponding CBF constraints into an online QP and establish guarantees for satisfaction of the counting tasks.

\begin{proposition} \label{prop:cbf}
    Consider the system in \eqref{eq:system}, a $C^1$ function $H : \mathbb{R}^{M n} \to \mathbb{R}$, and a locally Lipschitz extended class-$\mathcal{K}$ function $\alpha$.
    Let $\widetilde{\mathcal{D}} \subseteq \mathbb{R}^{M n}$ be an open set that contains the set $\widetilde{\mathcal{C}} = \left\{ \widetilde{\mathbf{x}} \in \mathbb{R}^{M n} : H (\widetilde{\mathbf{x}}) \ge 0 \right\} \subset \widetilde{\mathcal{D}}$.
    
    \noindent \textbf{(i)} Suppose that $H$ is a CBF as in Def.~\ref{def:cbf}. Then, any locally Lipschitz controller $\widetilde{\mathbf{u}} (\widetilde{\mathbf{x}}) = \begin{bmatrix} \mathbf{u}_1^\top & \dots & \mathbf{u}_M^\top \end{bmatrix}^\top$ satisfying 
    \begin{equation} \label{eq:cbf2}
        \sum_{i=1}^{M} \frac{\partial H}{\partial \mathbf{x}_i} (\widetilde{\mathbf{x}}) \left( \mathbf f_i(\mathbf{x}_i) + \mathbf G_i(\mathbf{x}_i) \, \mathbf u_i \right) \ge -\alpha (H (\widetilde{\mathbf{x}}))
    \end{equation}
    for all $\widetilde{\mathbf{x}} \in \widetilde{\mathcal{D}}$ renders $\widetilde{\mathcal{C}}$ forward invariant over the maximal interval of existence of each closed-loop solution. 

    \noindent \textbf{(ii)} Let $\widetilde{\mathcal{T}} \subseteq \mathbb R^{M n}$ be a closed set with $\widetilde{\mathcal{C}} \subseteq \widetilde{\mathcal{T}}$. Suppose there exists a class-$\mathcal{K}$ function $\gamma$ such that
    \begin{equation} \label{eq:cbf_distance_bound}
        \gamma \left( \| \widetilde{\mathbf{x}} \|_{\widetilde{\mathcal{T}}} \right) \le -H(\widetilde{\mathbf{x}})
    \end{equation}
    for all $\widetilde{\mathbf{x}} \notin \widetilde{\mathcal{T}}$. Then, any locally Lipschitz controller $\widetilde{\mathbf{u}} (\widetilde{\mathbf{x}})$ satisfying \eqref{eq:cbf2} for all $\widetilde{\mathbf{x}} \notin \widetilde{\mathcal{C}}$ renders $\widetilde{\mathcal{T}}$ attractive, provided that the closed-loop solutions are forward complete. That is, $ \lim_{t \to \infty} \| \widetilde{\mathbf{x}}(t) \|_{\widetilde{\mathcal{T}}} = 0$ for any $\widetilde{\mathbf{x}}(0) \in \mathbb{R}^{M n} \setminus \widetilde{\mathcal{T}}$.
\end{proposition}
\begin{proof}
    \textbf{(i)} Let $\widetilde{\mathbf{x}}(0) \in \widetilde{\mathcal{C}}$ and define $z(t) \coloneqq H(\widetilde{\mathbf{x}}(t))$. By \eqref{eq:cbf2}, we have $\dot z \ge -\alpha(z)$. Consider the comparison system $\dot w = -\alpha(w)$, with $w(0) = z(0) \ge 0$. Since $\alpha$ is extended class-$\mathcal{K}$ and locally Lipschitz, it holds that $w(t) \ge 0$, $t \in I \left( \widetilde{\mathbf{x}}(0) \right)$. By the comparison lemma \cite{khalil2002}, $z(t) \! \ge \! w(t) \! \ge  \!0$, $t \in I \left( \widetilde{\mathbf{x}}(0) \right)$.
    Hence, every closed-loop trajectory with $\widetilde{\mathbf{x}}(0) \in \widetilde{\mathcal{C}}$ remains in $\widetilde{\mathcal{C}}$ over $I \left( \widetilde{\mathbf{x}}(0) \right)$.

    \noindent \textbf{(ii)} Let $\widetilde{\mathbf{x}}(0) \notin \widetilde{\mathcal{T}}$. Then $H(\widetilde{\mathbf{x}}(0)) < 0$, since $\widetilde{\mathcal{C}} \subseteq \widetilde{\mathcal{T}}$. 
    Define $V(t) \coloneqq \max\{-H(\widetilde{\mathbf{x}}(t)), 0\}$ and a class-$\mathcal{K}$ function $\beta(s) \coloneqq -\alpha(-s)$.
    The function $V$ is locally absolutely continuous, and \eqref{eq:cbf2} gives $\dot V \le \alpha(-V) = -\beta(V) $ whenever $V > 0$. When $V = 0$, its derivative is zero almost everywhere, so the inequality holds almost everywhere along the trajectory. Since $\beta$ is locally Lipschitz and class-$\mathcal{K}$, comparing with $\dot w = -\beta(w)$, $w(0) = V(0) \ge 0$, 
     yields $0 \le V(t) \le w(t) \to 0$ as $t \to \infty$,
     provided that the closed-loop solution is forward complete. Moreover, $\gamma \left( \| \widetilde{\mathbf{x}}(t) \|_{\widetilde{\mathcal{T}}} \right) \le V(t)$, $ \forall t \in \mathbb{R}_{\ge 0}$. This follows from \eqref{eq:cbf_distance_bound} for $\widetilde{\mathbf{x}} \notin \widetilde{\mathcal{T}}$, while $\gamma \left( \| \widetilde{\mathbf{x}}(t) \|_{\widetilde{\mathcal{T}}} \right) = 0$ otherwise. Since $\gamma$ is class-$\mathcal{K}$ and $V(t) \to 0$ as $t \to \infty$, we have $\lim_{t \to \infty} \|\widetilde{\mathbf{x}}(t)\|_{\widetilde{\mathcal{T}}} = 0$, proving attractivity of $\widetilde{\mathcal{T}}$. 
\end{proof}

Consider the counting invariance and convergence tasks as in Defs.~\ref{def:counting_invariance}-\ref{def:counting_convergence}, associated with 
$m_\mathrm{inv}, m_\mathrm{conv} \in \{1, \dots, M\}$, and $C^1$ functions $h_\mathrm{inv}, h_\mathrm{conv}: \mathbb{R}^n \to \mathbb{R}$, defining $\mathcal{C}_\mathrm{inv} = \{ \mathbf{x} : h_\mathrm{inv} (\mathbf{x}) \ge 0 \}$ and $\mathcal{C}_\mathrm{conv} = \{ \mathbf{x} : h_\mathrm{conv} (\mathbf{x}) \ge 0 \}$, respectively. 

From \eqref{eq:lower_approximation1}, we obtain candidate CBFs $H_\mathrm{inv} \left( \widetilde{\mathbf{x}} \right) \coloneqq \ubar{h}_{(m_\mathrm{inv})} \left( \widetilde{\mathbf{x}} \right)$ and $H_{\mathrm{conv}} \left( \widetilde{\mathbf{x}} \right) \coloneqq \ubar{h}_{(m_{\mathrm{conv}})} \left( \widetilde{\mathbf{x}} \right)$. The corresponding sets associated with the CBFs are $\widetilde{\mathcal{C}}_{\mathrm{inv}} = \{\widetilde{\mathbf{x}}: H_{\mathrm{inv}}(\widetilde{\mathbf{x}}) \ge 0\}$, $\widetilde{\mathcal{C}}_{\mathrm{conv}} = \{\widetilde{\mathbf{x}}: H_{\mathrm{conv}}(\widetilde{\mathbf{x}}) \ge 0\}$. Since $H_{\mathrm{inv}}$ (resp. $H_{\mathrm{conv}}$) provides a lower approximation of the $m^\mathrm{th}$ largest value with respect to $C_\mathrm{inv}$ (resp. $C_\mathrm{conv}$), we get 
\begin{align}
    \widetilde{\mathcal{C}}_{\mathrm{inv}} &\subseteq \widetilde{\mathcal{T}}_\mathrm{inv} = \left\{\widetilde{\mathbf{x}}: \sum_{i=1}^{M} \mathbb{I}_{\mathcal{C}_{\mathrm{inv}}}(\mathbf{x}_i) \ge m_{\mathrm{inv}}\right\}, \label{eq:subset_inv} \\ 
    \widetilde{\mathcal{C}}_{\mathrm{conv}} &\subseteq \widetilde{\mathcal{T}}_\mathrm{conv} = \left\{\widetilde{\mathbf{x}}: \sum_{i=1}^{M} \mathbb{I}_{\mathcal{C}_{\mathrm{conv}}}(\mathbf{x}_i) \ge m_{\mathrm{conv}}\right\}. \label{eq:subset_conv}
\end{align}

To apply Prop.~\ref{prop:cbf}\textbf{(ii)} to counting convergence tasks, we must verify \eqref{eq:cbf_distance_bound} for the candidate CBF 
$H=H_{\mathrm{conv}}$ and $\widetilde{\mathcal{T}}=\widetilde{\mathcal{T}}_{\mathrm{conv}}$.
The following lemma derives this bound from a corresponding condition on $h_{\mathrm{conv}}$ that can be readily checked when choosing $h_{\mathrm{conv}}$.

\begin{lemma} \label{lemma}
    Suppose there exists a class-$\mathcal{K}$ function $\gamma_h$ such that 
    \begin{equation} \label{eq:lemma}
        \gamma_h (\|\mathbf{x}\|_{\mathcal{C}_\mathrm{conv}}) \le -h_\mathrm{conv} (\mathbf{x}) 
    \end{equation}
    for all $\mathbf{x} \notin \mathcal{C}_\mathrm{conv}$. Then \eqref{eq:cbf_distance_bound} holds with $H=H_{\mathrm{conv}}$,  $\widetilde{\mathcal{T}}=\widetilde{\mathcal{T}}_{\mathrm{conv}}$, and $\gamma(s)=\gamma_h(s/\sqrt{m_{\mathrm{conv}}})$.
    
\end{lemma}
\hspace{-2.4mm}\begin{proof}
    Let $h = h_\mathrm{conv}$, $m = m_\mathrm{conv}$, and $\mathbf{x}_{(1)}, \dots, \mathbf{x}_{(m)}$ denote the states corresponding to the $m$ largest values $h(\mathbf{x}_i)$. 
    For $\widetilde{\mathbf{x}} \notin \widetilde{\mathcal{T}}_\mathrm{conv}$, $h_{(m)}(\widetilde{\mathbf{x}}) < 0$. 
    By definition, we have $h \left( \mathbf{x}_{(i)} \right) \ge h_{(m)}(\widetilde{\mathbf{x}})$, $i=1, \dots, m$. Thus, by \eqref{eq:lemma} and  $\gamma_h(0)=0$, each agent~$i=1,\dots,m$ satisfies $\gamma_h (\|\mathbf{x}_{(i)}\|_{C_\mathrm{conv}}) \le \max \{ -h \left( \mathbf{x}_{(i)} \right), 0 \} \le -h_{(m)}(\widetilde{\mathbf{x}})$.
    Define $d_\mathrm{max} \coloneqq \max_{i=1, \dots, m} \| \mathbf{x}_{(i)} \|_{\mathcal{C}_\mathrm{conv}}$. Then,
    \begin{align*}
        \| \widetilde{\mathbf{x}} \|_{\widetilde{\mathcal{T}}_\mathrm{conv}} \!\le\!
        \sqrt{ \| \mathbf{x}_{(1)} \|_{\mathcal{C}_\mathrm{conv}}^2 \!+\! \dots \!+\! \| \mathbf{x}_{(m)} \|_{\mathcal{C}_\mathrm{conv}}^2 } \le 
        \sqrt{m} \, d_\mathrm{max}.
    \end{align*}
Monotonicity of $\gamma_h$ then gives 
\begin{align*}
        \gamma_h \left( \frac{\| \widetilde{\mathbf{x}} \|_{\widetilde{\mathcal{T}}_\mathrm{conv}}}{\sqrt{m}} \right) \le \gamma_h \left( d_\mathrm{max}  \right) \le -h_{(m)} (\widetilde{\mathbf{x}}).
\end{align*}
  The function $\gamma(s) \coloneqq \gamma_h (\nicefrac{s}{\sqrt{m}})$ is class-$\mathcal{K}$.
   Moreover, by \eqref{eq:approx_width1}, 
   $h_{(m)} (\widetilde{\mathbf{x}}) \ge H_\mathrm{conv}(\widetilde{\mathbf{x}})$, thus $\gamma \left( \|\widetilde{\mathbf{x}}\|_{\widetilde{\mathcal{T}}_\mathrm{conv}} \right) \le -H_{\mathrm{conv}} (\widetilde{\mathbf{x}})$, which completes the proof.
\end{proof}

We synthesize controllers for both counting tasks by solving a convex QP. At each state $\widetilde{\mathbf{x}}$, the optimal joint input $\widetilde{\mathbf{u}}^\star (\widetilde{\mathbf{x}}) = \begin{bmatrix} {\mathbf{u}_1^\star}^\top & \dots & {\mathbf{u}_M^\star}^\top \end{bmatrix}^\top$ is obtained by solving
\begin{subequations} \label{eq:qp}
    \begin{align}
        \min_{\widetilde{\mathbf{u}}, \delta_{\mathrm{inv}}, \delta_{\mathrm{conv}}} \quad&
        \frac12 \left\| \widetilde{\mathbf{u}} -\widetilde{\mathbf{u}}_{\mathrm{nom}} \left( \widetilde{\mathbf{x}} \right) \right\|^2 + \frac{p_{\mathrm{inv}}}{2} \, \delta_{\mathrm{inv}}^2 + \frac{p_{\mathrm{conv}}}{2} \, \delta_{\mathrm{conv}}^2 
        \notag \\
        \text{s.t.} \quad& \Psi_{\mathrm{inv}} \left( \widetilde{\mathbf{x}}, \widetilde{\mathbf{u}} \right) \ge -\delta_{\mathrm{inv}}
        \label{eq:counting_qp_inv} \\
        &
        \Psi_{\mathrm{conv}} \left( \widetilde{\mathbf{x}}, \widetilde{\mathbf{u}} \right) \ge -\delta_{\mathrm{conv}}
        \label{eq:counting_qp_conv}
    \end{align}
\end{subequations}
where $\widetilde{\mathbf{u}}_{\mathrm{nom}} \left( \widetilde{\mathbf{x}} \right) \in \mathbb{R}^{M p}$ is a locally Lipschitz nominal controller encoding the individual objectives of the agents, $\delta_{\mathrm{inv}}, \delta_{\mathrm{conv}} \in \mathbb{R}$ are slack variables, and $p_{\mathrm{inv}}, p_{\mathrm{conv}} > 0$ are their respective penalty weights. For $j \in \{ \mathrm{inv}, \mathrm{conv} \}$, the CBF constraints are defined by
\[
    \Psi_j (\widetilde{\mathbf{x}}, \widetilde{\mathbf{u}}) = \sum_{i=1}^{M} \frac{\partial H_j}{\partial\mathbf{x}_i} (\widetilde{\mathbf{x}}) \left( \mathbf{f}_i (\mathbf{x}_i) + \mathbf{G}_i (\mathbf{x}_i) \, \mathbf{u}_i \right) + \alpha(H_j(\widetilde{\mathbf{x}})),
\]
with $\alpha$ being locally Lipschitz and extended class-$\mathcal{K}$. 
The gradient of $H_j(\widetilde{\mathbf{x}}) = \ubar{h}_{(m_j)} (\widetilde{\mathbf{x}})$ is efficiently evaluated through the backward recursion in \eqref{eq:derivative1}. The closed-loop trajectories of \eqref{eq:system} are obtained by applying $\mathbf{u}_i^\star (\widetilde{\mathbf{x}})$, $i \in \mathcal{I}$.
\begin{remark} \label{rem:relaxation}
    The QP in \eqref{eq:qp} allows the two counting tasks to be enforced separately, relaxed or prioritized. In particular, setting $\delta_{\mathrm{inv}} = 0$ (resp. $\delta_{\mathrm{conv}} = 0$) enforces the counting invariance (resp. convergence) constraint without relaxation. In this case, satisfaction guarantees can be established for the prioritized task, whenever the QP admits a solution.
\end{remark}

The following theorem establishes sufficient conditions under which $\widetilde{\mathbf{u}}^\star (\widetilde{\mathbf{x}})$ from \eqref{eq:qp} guarantees satisfaction of the counting invariance task (with $\delta_\mathrm{inv} = 0$) and the counting convergence task (with $\delta_\mathrm{conv} = 0$).

\begin{theorem} \label{thm:main}
    Consider the system in \eqref{eq:system} and the counting tasks in Def.~\ref{def:counting_invariance}-\ref{def:counting_convergence}, with $m_j$ and functions $h_j$ defining $\mathcal{C}_j$,  $j \in \{\mathrm{inv},\mathrm{conv}\}$. Let $H_j$ be constructed as in \eqref{eq:lower_approximation1}, with its value and gradient evaluated using Alg.~\ref{alg:recursion}. Assume that $h_\mathrm{inv}$ and $h_\mathrm{conv}$ have locally Lipschitz gradients. Consider the controller obtained from the QP in \eqref{eq:qp}, and assume that the resulting closed-loop solutions are forward complete.

    \noindent \textbf{(i) Counting invariance.} Consider $\widetilde{\mathcal{C}}_{\mathrm{inv}} \!=\! \{\widetilde{\mathbf{x}}:\! H_{\mathrm{inv}}(\widetilde{\mathbf{x}}) \!\ge\! 0\}$. Let $\widetilde{\mathcal{D}}_\mathrm{inv} \supset \widetilde{\mathcal{C}}_\mathrm{inv}$ be an open set such that, for every $\widetilde{\mathbf{x}} \in \widetilde{\mathcal{D}}_\mathrm{inv}$, there exists $i \in \mathcal{I}$  such that $\frac{\partial H_{\mathrm{inv}}}{\partial \mathbf{x}_i} (\widetilde{\mathbf{x}}) \mathbf{G}_i(\mathbf{x}_i) \ne \mathbf{0}^{\top}.$
    Then the QP controller in \eqref{eq:qp}  with $\delta_{\mathrm{inv}}=0$ satisfies the counting invariance task for every $\widetilde{\mathbf{x}}(0)\in\widetilde{\mathcal{C}}_\mathrm{inv}$.   

    \noindent \textbf{(ii) Counting convergence.} 
    Consider $\widetilde{\mathcal{C}}_{\mathrm{conv}} \!=\! \{\widetilde{\mathbf{x}}:\! H_{\mathrm{conv}}(\widetilde{\mathbf{x}}) \!\ge\! 0\}$.
    Suppose that for every $\widetilde{\mathbf{x}} \notin \widetilde{\mathcal{C}}_\mathrm{conv}$, there exists $i \in \mathcal{I}$ such that $      \frac{\partial H_{\mathrm{conv}}}{\partial\mathbf{x}_i}(\widetilde{\mathbf{x}}) \mathbf{G}_i(\mathbf{x}_i)\ne\mathbf{0}^{\top}.$
    Assume that \eqref{eq:lemma} holds for all $\mathbf{x} \notin \mathcal{C}_{\mathrm{conv}}$ with a class-$\mathcal{K}$
    function $\gamma_h$. 
    Then, the QP controller in \eqref{eq:qp} with $\delta_{\mathrm{conv}} = 0$  satisfies the counting convergence task for every $\widetilde{\mathbf{x}}(0) \notin \widetilde{\mathcal{T}}_\mathrm{conv}$.
        
\end{theorem}
\begin{proof}
    \noindent \textbf{(i)} Since the gradient of $h_\mathrm{inv}$ is locally Lipschitz, the gradient of $H_\mathrm{inv} = \ubar{h}_{(m_\mathrm{inv})}$ is locally Lipschitz, as follows from \eqref{eq:derivative1}. By the nonvanishing condition $  \frac{\partial H_{\mathrm{inv}}}{\partial\mathbf{x}_i}(\widetilde{\mathbf{x}}) \mathbf{G}_i(\mathbf{x}_i)\ne\mathbf{0}^{\top}$, the active constraints of the QP in \eqref{eq:qp} with $\delta_{\mathrm{inv}} = 0$ are linearly independent on $\widetilde{\mathcal{D}}_\mathrm{inv}$. Note that without input bounds, and by choosing the slack variable $\delta_{\mathrm{conv}}$ sufficiently large, the inequality in \eqref{eq:qp} is always feasible for all $\widetilde{\mathbf{x}} \in \widetilde{\mathcal{D}}_\mathrm{inv}$. 
    Together with strong convexity of the objective and local Lipschitz continuity of $\alpha$, $\widetilde{\mathbf u}_{\mathrm{nom}}$, $\mathbf{f}_i$ and $\mathbf{G}_i$, $i \in \mathcal{I}$, this implies that the unique optimizer $\widetilde{\mathbf{u}}^\star (\widetilde{\mathbf{x}})$ is locally Lipschitz on $\widetilde{\mathcal{D}}_\mathrm{inv}$. According to Def.~\ref{def:cbf}, we get that the $H_\mathrm{inv}$ constructed in \eqref{eq:lower_approximation1} is indeed a CBF. Then, applying Prop.~\ref{prop:cbf}\textbf{(i)}, together with forward completeness, we get that
    $\widetilde{\mathcal{C}}_\mathrm{inv}$ is forward invariant $\forall t \in \mathbb{R}_{\ge 0}$. Recalling that $\widetilde{\mathcal{C}}_\mathrm{inv} \subseteq \widetilde{\mathcal{T}}_\mathrm{inv}$ by \eqref{eq:subset_inv}, the counting invariance task is satisfied for every $\widetilde{\mathbf{x}}(0)\in\widetilde{\mathcal{C}}_\mathrm{inv}$.   
    
    \noindent \textbf{(ii)} Similarly, the nonvanishing condition ensures linear independence of the active constraints of the QP in \eqref{eq:qp} with $\delta_{\mathrm{conv}} = 0$. Thus, $\widetilde{\mathbf{u}}^\star (\widetilde{\mathbf{x}})$ is locally Lipschitz on $\mathbb{R}^{M n} \setminus \widetilde{\mathcal{C}}_\mathrm{conv}$. Moreover, with the QP being feasible for all $\widetilde{\mathbf{x}} \notin \widetilde{\mathcal{C}}_\mathrm{conv}$, we get that $H_\mathrm{conv}$ constructed in \eqref{eq:lower_approximation1} satisfies \eqref{eq:cbf2} for all $\widetilde{\mathbf{x}} \notin \widetilde{\mathcal{C}}_\mathrm{conv}$. 
    Since \eqref{eq:lemma} holds for $\mathbf{x} \notin \mathcal{C}_{\mathrm{conv}}$, by Lemma~\ref{lemma}, \eqref{eq:cbf_distance_bound} holds for all $\widetilde{\mathbf{x}} \notin \widetilde{\mathcal{T}}_\mathrm{conv}$. 
    Recalling that $\widetilde{\mathcal{C}}_\mathrm{conv} \subseteq\widetilde{\mathcal{T}}_\mathrm{conv}$ by~\eqref{eq:subset_conv}. 
    By applying Prop.~\ref{prop:cbf}\textbf{(ii)}, together with the forward completeness, we get that $\widetilde{\mathcal{T}}_\mathrm{conv}$ is attractive. Hence, the counting convergence task is satisfied for every $\widetilde{\mathbf{x}}(0) \notin \widetilde{\mathcal{T}}_\mathrm{conv}$. 
\end{proof}

\section{Numerical Results} \label{sec:numerical}

In this section, we evaluate the scalability of the proposed recursion-based method and illustrate controller synthesis for counting invariance and counting convergence tasks through a multi-drone emergency response case study. Simulations were run in MATLAB R2025a on a computer equipped with an AMD Ryzen 7 PRO 8845HS processor and 32 GB of RAM.

\subsection{Computational performance and scalability}

\begin{figure}[t]
    \begin{center}
        \includegraphics[width=0.8\columnwidth]{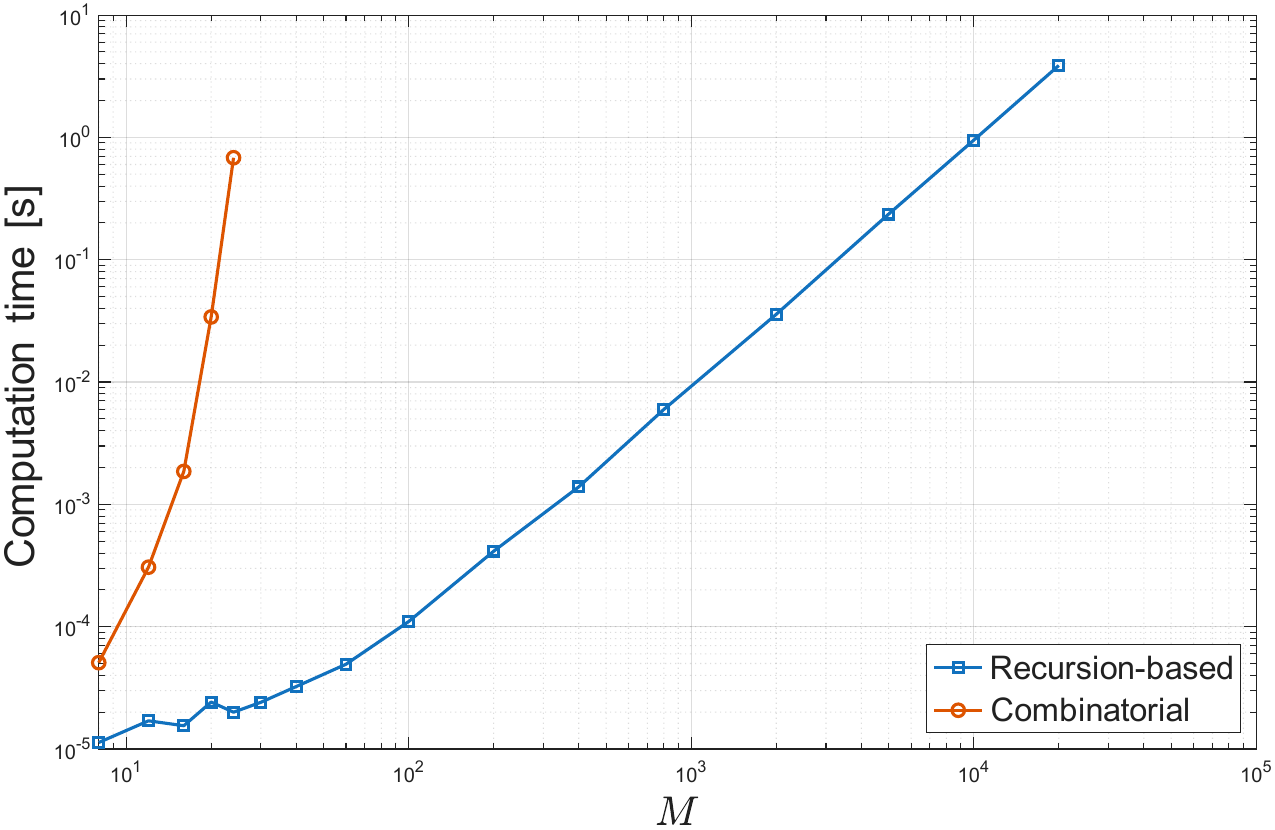}
        \caption{Experiment~(i): Computation times of the recursion-based and combinatorial methods as $M$ increases, with $\nicefrac{m}{M} \approx 0.4$.} 
        \label{fig:comparison1}
    \end{center}
        \vspace{-0.5cm}
\end{figure}

\begin{figure}[t!]
    \begin{center}
        \includegraphics[width=0.8\columnwidth]{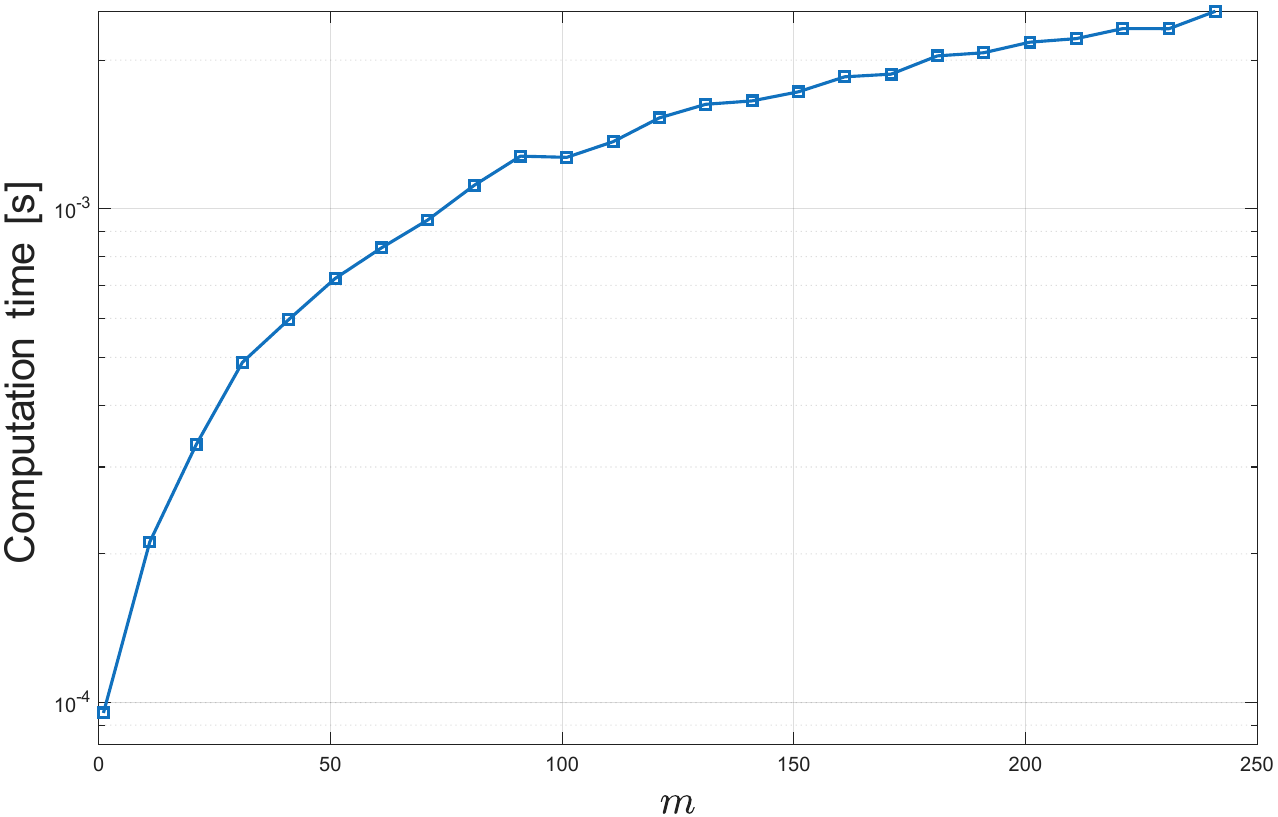}
        \caption{Experiment~(ii): Computation times of the recursion-based method as $m$ increases, with $M=500$.} 
        \label{fig:comparison2}
    \end{center}
    \vspace{-0.7cm}
\end{figure}

Let $M$ denote the total number of agents and $m$ the required agent count.
We evaluate the computation times required to calculate the CBF and its gradient using the recursion-based method in Sec.~\ref{sec:cbf}. For comparison, we consider a na{\"i}ve combinatorial method that explicitly enumerates all subsets of cardinalities $m$ and $m-1$, computes the LogSumExp approximations $\hat{\sigma}_m$ and $\hat{\sigma}_{m-1}$ from \eqref{eq:hat_sum_k_largest}, and evaluates the candidate CBF $\ubar{h}_{(m)}$ in \eqref{eq:underapprox_recursion} and its gradient. Both methods yield the same CBF and gradient, but the recursion-based method avoids combinatorial computational complexity by reusing intermediate quantities. 
Two experiments are conducted in MATLAB: (i) varying $M$ from 8 to 20000, with $m / M \approx 0.4$, and (ii) varying $m$ from 1 to 241, with $M = 500$. For each experiment, the task function values $h_i$  are randomly generated, and the median computation times over 15 evaluations are reported. We choose
\begin{equation*}
    \tau = \frac{W}{ \ln \binom{M}{m} + \ln \binom{M}{m-1} }, \quad W = 0.1,
\end{equation*}
to ensure $0 \!\le\! h_{(m)} \!- \!\ubar{h}_{(m)} \!\le\! W$, according to \eqref{eq:approx_width1}.
The results in Fig.~\ref{fig:comparison1}-\ref{fig:comparison2} are consistent with the complexity analysis in Sec.~\ref{sec:complexity}. 
In Fig.~\ref{fig:comparison1}, computation time of the proposed approach scales approximately quadratically with $M$ at a fixed ratio $\nicefrac{m}{M}$, remaining below $10~\mathrm{ms}$ even at $M\!=\!800$. 
In contrast, combinatorial enumeration becomes prohibitive for online computation with only $M \!=\! 20$ agents.
In Fig.~\ref{fig:comparison2}, computation time scales approximately linearly with $m$ for $M\!=\!500$, remaining below $3~\mathrm{ms}$ for all tested values of $m$. 

These computation times quantify the cost of evaluating the CBF and its gradient at each sampling instant before solving the QP in \eqref{eq:qp}. 
The results demonstrate the potential of the proposed method for real-time control of hundreds of agents, even when the task requires a large fraction of the team to satisfy the prescribed condition.

\subsection{Case study}

\begin{figure}
    \centering
    \begin{subfigure}[b]{0.45\columnwidth}
        \centering
        \includegraphics[width=\textwidth]{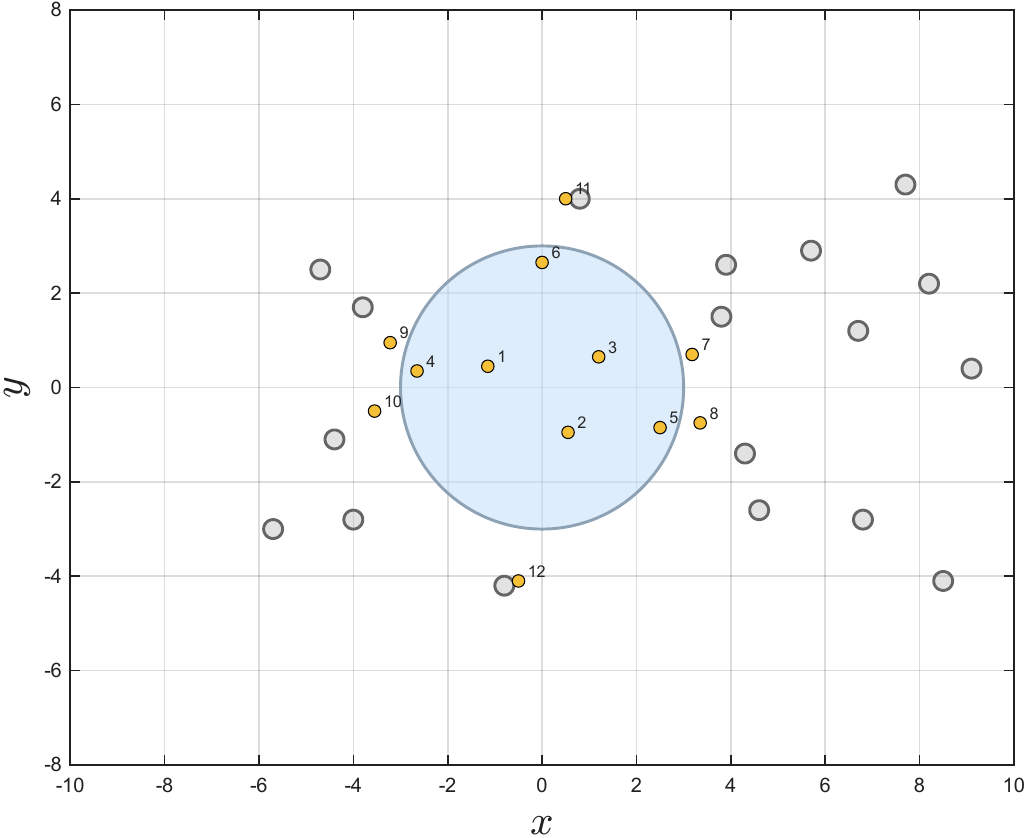}
        \vspace{-6mm}
        \caption{t=0}
        \label{fig:t=0}
    \end{subfigure}
    \hfill
    \begin{subfigure}[b]{0.45\columnwidth}  
        \centering 
        \includegraphics[width=\textwidth]{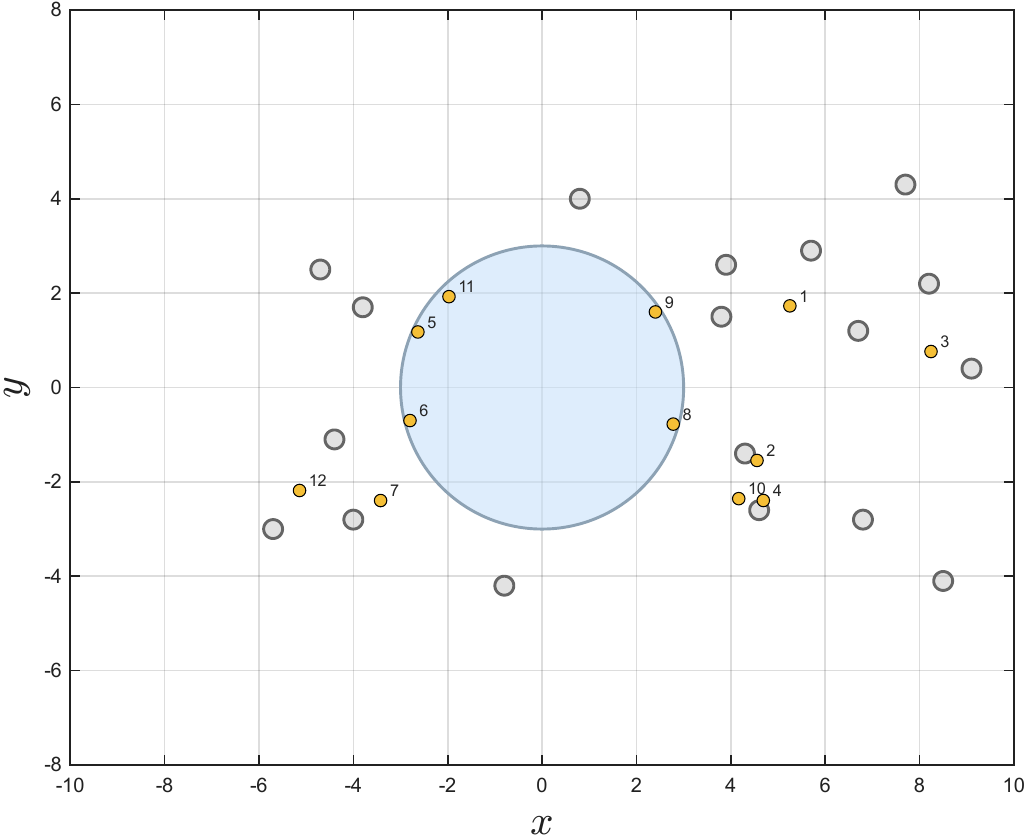}
        \vspace{-6mm}
        \caption{t=3.5}
        \label{fig:t=3.5}
    \end{subfigure}
    \begin{subfigure}[b]{0.45\columnwidth}   
        \centering 
        \includegraphics[width=\textwidth]{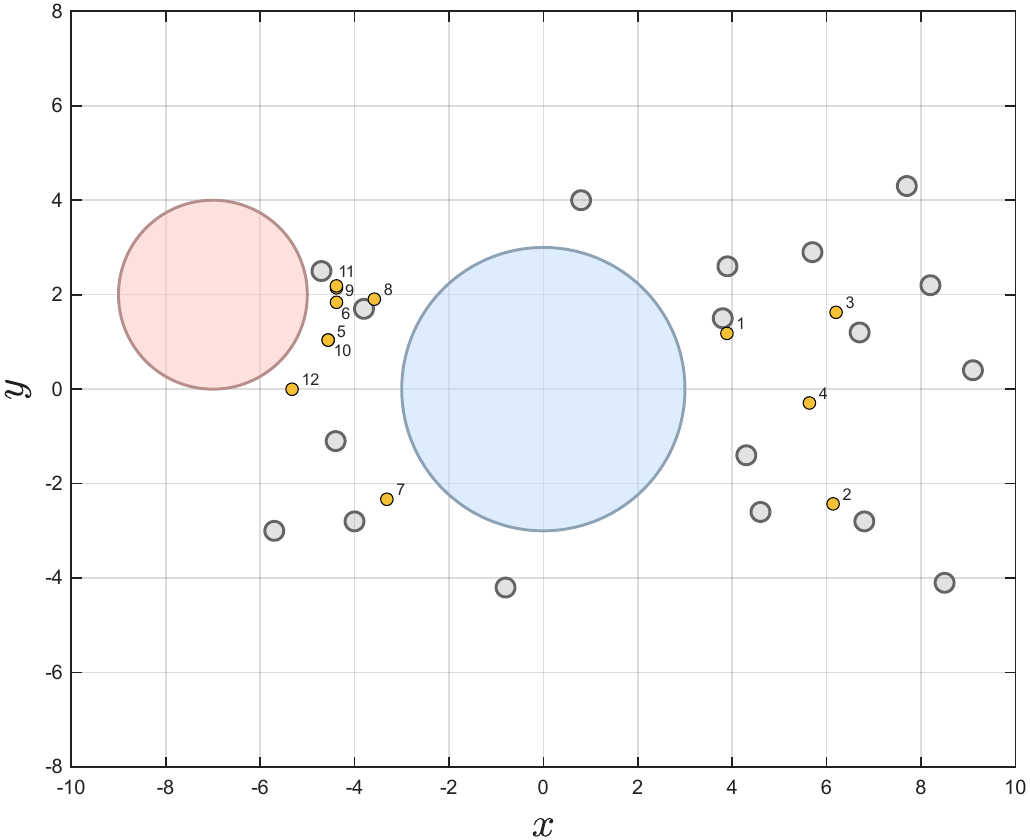}
        \vspace{-6mm}
        \caption{t=8.3}
        \label{fig:t=8.3}
    \end{subfigure}
    \hfill
    \begin{subfigure}[b]{0.45\columnwidth}   
        \centering 
        \includegraphics[width=\textwidth]{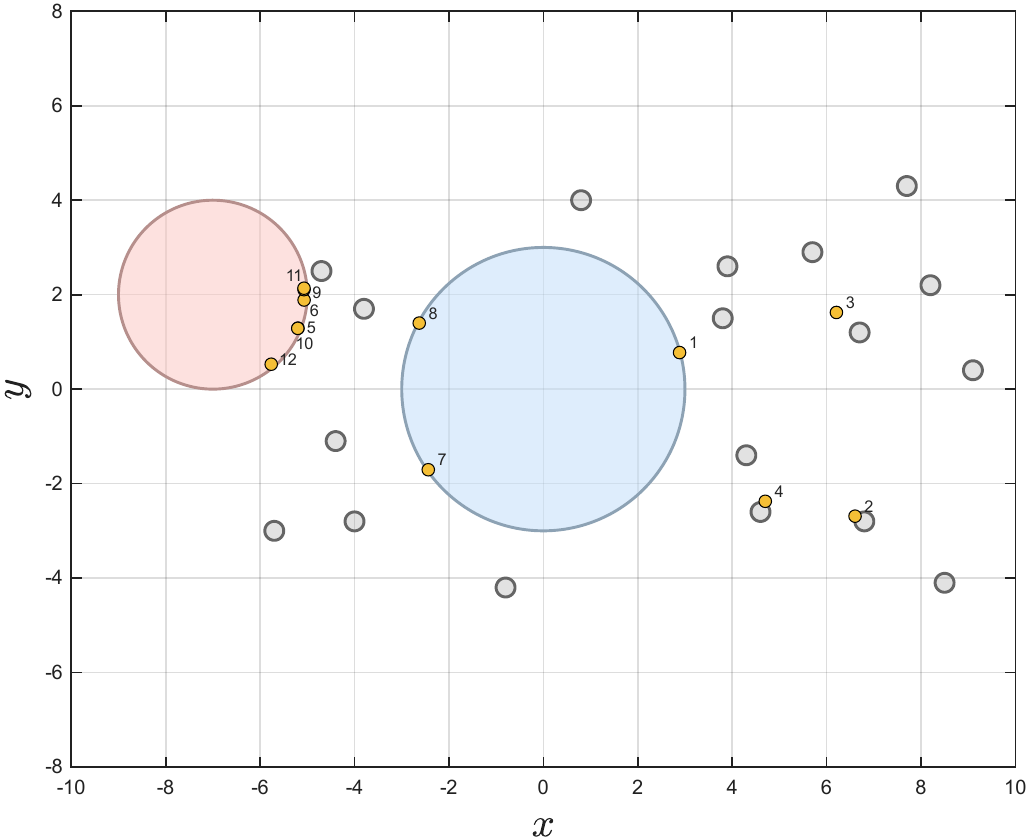}  
        \vspace{-6mm}
        \caption{t=15}
        \label{fig:t=15}
    \end{subfigure}
    \caption{Drone positions at \textbf{(a)} $t=0$, \textbf{(b)} $t=3.5$ (before the aftershock), and \textbf{(c)} $t=8.3$, \textbf{(d)} $t=15$ (after). Drones (in yellow), villages (in gray), the standby region (in blue), and the emergency region (in red) are shown.}
    \label{fig:positions}
     \vspace{-0.5cm}
\end{figure}

We illustrate the proposed framework on a multi-drone emergency response scenario. A fleet of $M=12$ drones delivers supplies between villages following an earthquake, with each drone assigned two villages to visit alternately. The emergency response protocol requires at least $m_\mathrm{inv} = 5$ drones to remain in the standby region 
\[
    \mathcal{C}_\mathrm{inv} = \left\{ \mathbf{x} \in \mathbb{R}^2 : h_\mathrm{inv}(\mathbf{x}) = 9 - \|\mathbf{x}\|^2 \ge 0 \right\},
\]    
maintaining a reserve in case of potential aftershocks. This defines a counting invariance task as in Def.~\ref{def:counting_invariance}. Fig.~\ref{fig:positions} shows the drones in yellow, the villages in gray, and the standby region in blue.

At time $t=5$, an aftershock activates a counting convergence task as in Def.~\ref{def:counting_convergence} requiring at least $m_\mathrm{conv} = 6$ drones to converge to the emergency region 
\[
    \mathcal{C}_\mathrm{conv} =\left\{ \mathbf{x} \in \mathbb{R}^2 : h_\mathrm{conv}(\mathbf{x}) = 4 -\!\left\| \mathbf{x} - \begin{bmatrix} -7 \!&\! 2 \end{bmatrix}^\top \right\|^2 \ge 0 \right \}, 
\]
shown in red in Fig.~\ref{fig:positions}. 
The standby invariance task is reduced to $m_\mathrm{inv}' = 3$ drones, and its CBF constraint is relaxed using a slack variable $\delta_\mathrm{inv} \ge 0$ with penalty weight $p_\mathrm{inv} = 1000$ as in \eqref{eq:qp}. The emergency convergence task is enforced without relaxation  ($\delta_\mathrm{conv} = 0$), prioritizing the emergency response task over the standby invariance task as described in Remark~\ref{rem:relaxation}.

Each drone follows single-integrator dynamics, i.e.,
\begin{equation*}
    \dot{\mathbf{x}}_i = \mathbf{u}_i, \quad i \in \mathcal{I},
\end{equation*}
where $\mathbf{x}_i, \mathbf{u}_i \in \mathbb{R}^2$ denote its position and velocity, respectively. Both $h_\mathrm{inv}$ and $h_\mathrm{conv}$ have locally Lipschitz gradients. Moreover, $h_\mathrm{conv}$ satisfies \eqref{eq:lemma}, by choosing $\gamma_h(s) = s$.

Following Sec.~\ref{sec:cbf}, we construct CBFs for the counting tasks. For $t \in [0, 5)$, we construct $H_\mathrm{inv}$ from $h_\mathrm{inv}$ and $m_\mathrm{inv} = 5$. For $t \in [5, 15]$, we construct $H_\mathrm{conv}$ and $H_\mathrm{inv}'$ from $h_\mathrm{conv}$, $m_\mathrm{conv} = 6$, and $h_\mathrm{inv}$, $m_\mathrm{inv}' = 3$, respectively. 
We evaluate their values and gradients using the forward and backward recursions in Alg.~\ref{alg:recursion}.
We choose $\tau \approx 0.0074$, ensuring that the worst-case approximation error is at most 0.1 for all three counting tasks,  according to \eqref{eq:approx_width1}. The resulting CBF constraints are incorporated into the QP in \eqref{eq:qp} for controller synthesis.

The individual delivery objective of each drone~$i$, $i \in \mathcal{I}$, is encoded by the nominal input 
\[
    \mathbf{u}_{i, \mathrm{nom}} = - c_1 \!\cdot\! c_2 \frac{\|\mathbf{x}_i - \mathbf{r}_i \|^2}{2 \left( 1 + c_1 \|\mathbf{x}_i - \mathbf{r}_i \|^2 \right)} (\mathbf{x}_i - \mathbf{r}_i),
\]
where $\mathbf{r}_i$ is the center of the current target village, $c_1 = 10^5$ and $c_2 = 5$. Whenever a drone reaches within distance $0.2$ of its current target village's center, its target is updated to the other assigned village.

\begin{figure}[tbh]
    \begin{center}
        \includegraphics[width=0.9\columnwidth]{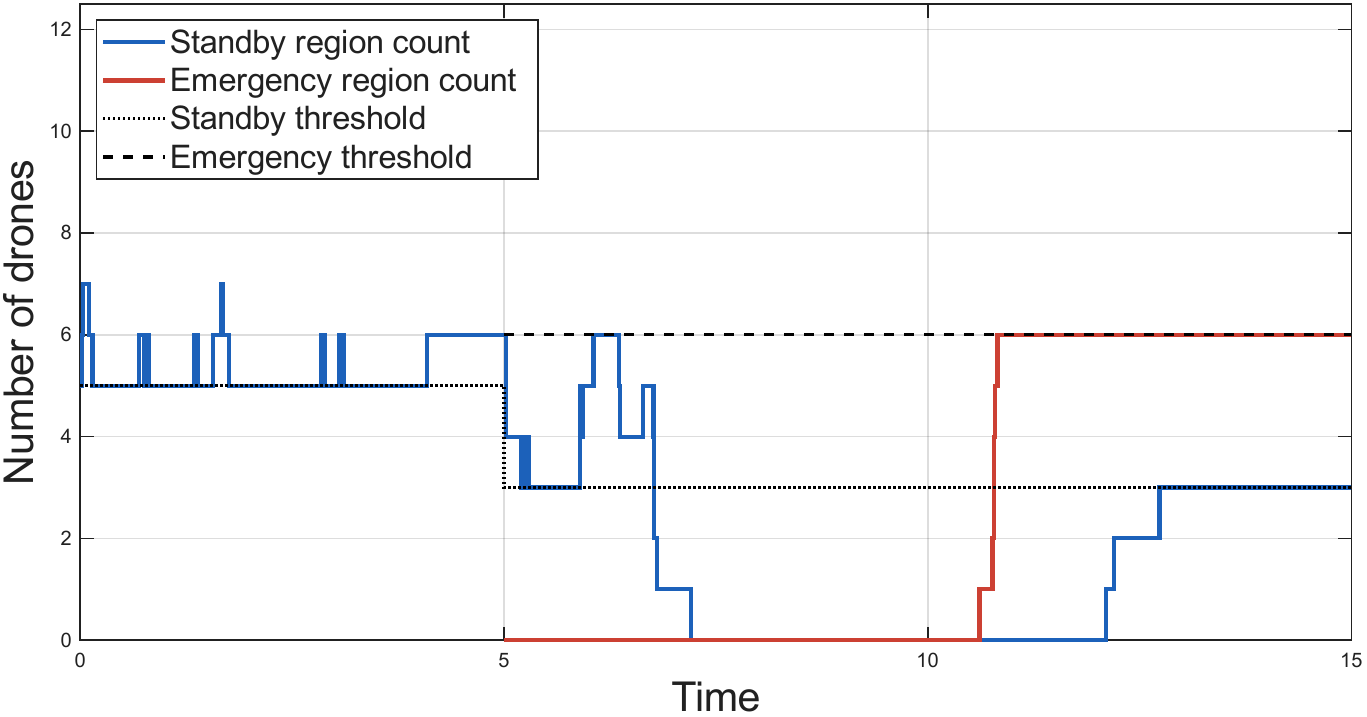}
        \caption{Drone counts and desired thresholds for the standby invariance and emergency convergence tasks.} 
        \label{fig:counts}
    \end{center}
     \vspace{-0.5cm}
\end{figure}

\begin{figure}[tbh]
    \begin{center}
        \includegraphics[width=0.9\columnwidth]{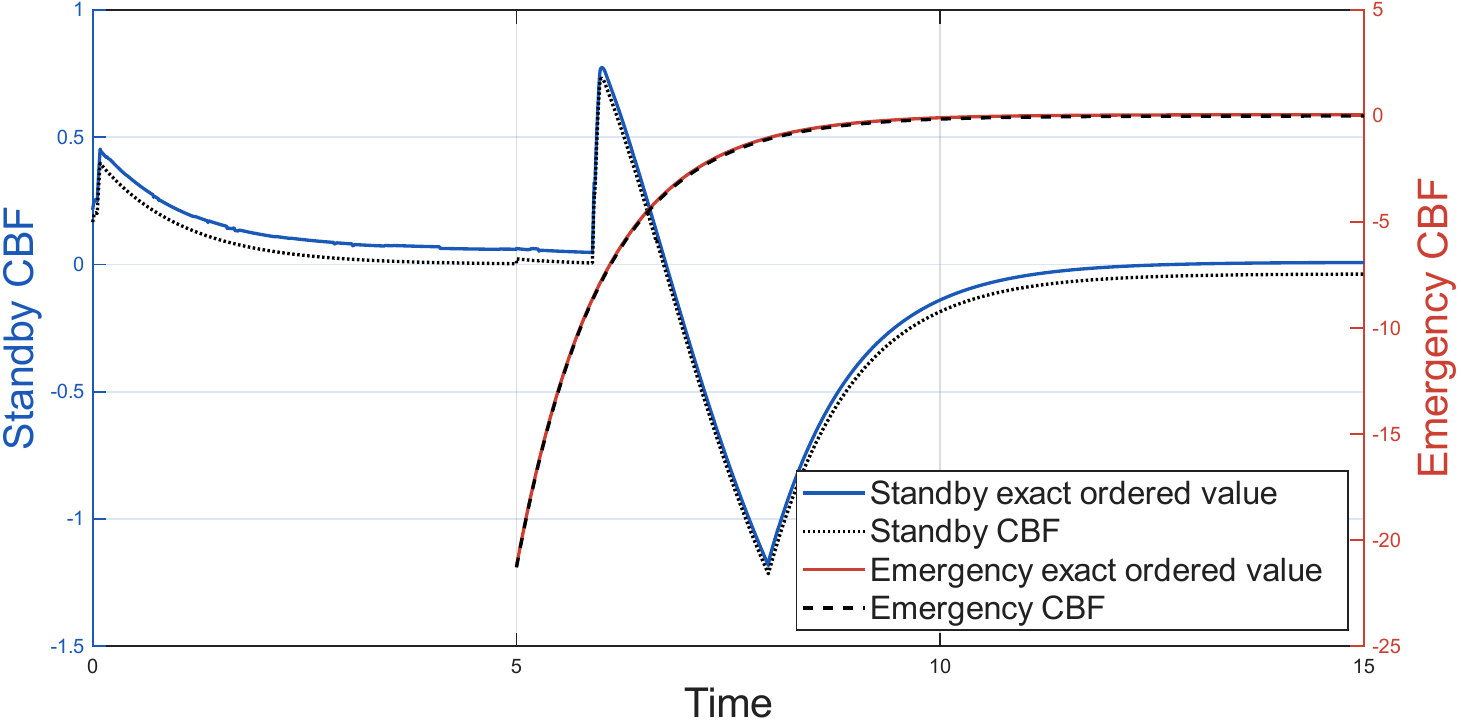}
        \caption{CBFs and exact ordered values for the standby invariance and emergency convergence tasks.} 
        \label{fig:barriers}
    \end{center}
     \vspace{-0.7cm}
\end{figure}

Fig.~\ref{fig:positions}\textbf{(a)}-\textbf{(d)} show snapshots of the fleet at initial deployment ($t=0$), before the aftershock ($t=3.5$), after the aftershock but before convergence ($t=8.3$) and after convergence ($t=15$), respectively. 
Fig.~\ref{fig:counts} shows the numbers of drones in the standby and emergency regions over time, together with the desired thresholds, i.e., $m_\mathrm{inv}=5$ for $t \in [0,5)$, and $m_\mathrm{conv}=6$, $m_\mathrm{inv}'=3$ for $t \in [5,15]$. The standby invariance task is satisfied throughout $[0,5)$, while the emergency region count first reaches six at $t=10.82$, corresponding to a response time of $5.82$ after the aftershock. 
Fig.~\ref{fig:barriers} relates these counts to the exact ordered values and the corresponding CBFs: at least $m$ drones belong to a region if and only if $h_{(m)} \ge 0$, while the CBF $H$ is a lower approximation of $h_{(m)}$. At $t=15$, both counting requirements hold, as confirmed by Fig.~\ref{fig:positions}-\ref{fig:barriers}. These results demonstrate that the proposed framework enables controller synthesis for both counting invariance and counting convergence tasks, while accommodating individual objectives.

\section{Conclusion} \label{sec:conclusion}

This paper proposes an online CBF-based control synthesis framework for counting invariance and counting convergence tasks in continuous-time multi-agent systems. A recursion-based method enables efficient construction and evaluation of differentiable CBFs without explicit task allocation or enumeration of agent subteams. The framework provides theoretical guarantees of counting-task satisfaction and is validated through a multi-drone emergency-response case study.
Future work includes extensions to more general cLTL specifications and higher-order dynamics through high-order CBFs \cite{xiao2021}, experimental validation, and distributed implementations.

\vspace{-0.3cm}

\bibliographystyle{IEEEtran}
\bibliography{ref}

\end{document}